\documentclass[aps,twocolumn,,superscriptaddress,showpacs,showkeys,amsmath,amssymb,floatfix]{revtex4}
\usepackage{graphicx}
\usepackage{epsfig}
\usepackage{dcolumn}
\usepackage{bm}
\usepackage{amssymb}
\usepackage{dsfont}
\usepackage{amsmath}
\usepackage{subfigure}
\newcommand{\be}{\begin{equation}}
\newcommand{\ee}{\end{equation}}
\newcommand{\bea}{\begin{eqnarray}}
\newcommand{\eea}{\end{eqnarray}}

\newcommand{\tC}{{\tilde C}}

\newcommand{\der}{\partial}

\newcommand{\ba}{\begin{array}}
\newcommand{\ea}{\end{array}}

\newcommand{\nn}{\nonumber}

\newcommand{\uast}{\stackrel{\ast}{u}}

\begin{document}

\title{Exact knot solutions in a generalized Skyrme-Faddeev model}
\bigskip
 \author{L. P. Zou}
\affiliation{Institute of Modern Physics, Chinese Academy of Sciences,
 Lanzhou 730000, China}
 \affiliation{
School of Nuclear Science and Technology, Lanzhou University, Lanzhou 730000, China}
\affiliation{
University of Chinese Academy of Sciences, Beijing 100049, China}
 \author{P. M. Zhang}
\affiliation{Institute of Modern Physics, Chinese Academy of Sciences,
 Lanzhou 730000, China}
  \affiliation{Research Center for Hadron and CSR Physics,
 Lanzhou University, Lanzhou 730000, China}
\author{D. G. Pak}
\affiliation{Institute of Modern Physics, Chinese Academy of Sciences,
 Lanzhou 730000, China}
\affiliation{Lab. of Few Nucleon Systems,
 Institute for Nuclear Physics, Ulugbek, 100214, Uzbekistan}

\begin{abstract}
We propose a generalized Skyrme-Faddeev type theory
with an additional scalar field. In a special case of model
parameters one has a theory which admits exact knot solutions
given by a class of exact toroidal solitons from
Aratyn-Ferreira-Zimerman (AFZ) integrable $CP^1$ model.
In a general case the theory admits an exact knot
solution for a unit Hopf charge. For higher Hopf charges
we perform numeric analysis of the solutions
and obtain estimates for the knot energies
using energy minimization procedure based on ansatz with
AFZ field configurations and with rational functions.
We show that AFZ configurations provide a better
approximate solutions. The corresponding knot energies
are in a good agreement with a standard law for the
low energy bound, $E_H\simeq Q_H^{3/4}$.
\end{abstract}
\vspace{0.3cm}
\maketitle

\vspace{2mm}

Since invention of the original Skyrme theory \cite{skyrme,zahed}
a great development towards its generalizations and
applications in description of baryons has been made
\cite{jackson,atkins,skyrmereview}.
Topological solitons in that theory, so called Skyrmions,
have been constructed numerically for a wide range of
baryon numbers \cite{battye,manton2012}.
The Skyrme-Faddeev theory \cite{FNnature,battyeprl}
represents another kind of effective theories
which supposed to be induced from
quantum chromodynamics (QCD) in low energy regime.
The theory possesses topological knot solitons
classified by the homotopy group $\pi_3(CP^1)=Z$,
i.e., by the topological Hopf charge. It is proposed
to interpret such knot solutions as color electric and color
magnetic glueball states
\cite{FNnature,choprl01,niemi00,choplb05,chozhang}.
One should notice, that derivation of a strict expression for
a low energy effective action from the basics of QCD is an
extremely difficult problem \cite{choleepak02,chopakkor01,gies}.
So far there exists a number of various extended Skyrme-Faddeev
models where some exact and numeric knot and vortex solutions have been found
\cite{ferr2012,ferr2011,ferr2009,shi}.

In the present paper we consider an extended Skyrme-Faddeev
model with an additional scalar field. In
 general the low energy effective action of QCD
 contains scalar fields \cite{choprl01} which originate
 from unknown coefficient correlation functions in the effective action
 and play roles of order parameters in the effective theory
\cite{choleepak02}. We consider a minimal extension of the
Skyrme-Faddeev theory to find out a model which is similar
to the Skyrme theory, and, at the same time, inherits
properties of integrable $CP^1$ models.

So far, all known topological solutions in $(3+1)$
Skyrme type models are obtained
numerically except the original Skyrmion soliton \cite{skyrme}.
Some exact vortex solutions have been obtained
in a generalized Skyrme-Faddeev theory
for a special set of model parameters which imply the existence of
integrable submodel in the theory \cite{ferr2012,ferr2011,ferr2009}.
A family of exact analytic knot solutions with toroidal configuration
have been constructed in special integrable $CP^1$ models
\cite{nicole, AFZ}. Unfortunately, physical
applications of such solutions in QCD remain unclear.
We propose a Skyrme-Faddeev-type model
which admits exact knot solutions and which can
give some hints towards constructing a more realistic
low energy QCD effective theory.

Let us consider an extended Skyrme-Faddeev model
defined by the following Lagrangian
\bea
{\cal L}(\phi, \hat n)&=&-\mu^2 (\der_\mu \phi)^2-\dfrac{\beta}{4} \phi^2 (\der_\mu \hat n)^2-
\dfrac{\nu}{32} \dfrac{1}{\phi^2} H_{\mu\nu}^2-\nn \\
           && \dfrac{\xi}{32}\dfrac{1}{\phi^2}(\der_\mu \hat n)^4,   \label{Skymod}
\eea
where $\hat n$ is a unit triplet field
in adjoint representation of $SU(2)$ color group,
i.e., $CP^1$ field,
 $\phi$ is a real scalar field, and $H_{\mu\nu}$ is a
magnetic field defined as follows
\bea
H_{\mu\nu}&=&\epsilon^{abc} \hat n^a \der_\mu \hat n^b \der_\nu \hat n^c,
\eea
$\mu, \beta,\nu,\xi$ are model parameters.
The first two terms in (\ref{Skymod}) coincide
with respective first two terms in the original Skyrme Lagrangian
after proper changing variable for the scalar field.
The last term in the Lagrangian appears in the one-loop
effective action of standard QCD as it was shown in \cite{choleepak02,gies}.
The topological content of the theory is determined by
the $CP^1$ field $\hat n$ realizing the Hopf mapping.
One has two non-trivial homotopy groups,
$\pi_2(CP^1)=Z$ and $\pi_3(CP^1)=Z$,
which classify all non-equivalent
topological classes of $\hat n$
by the monopole and Hopf charges respectively.
Like Skyrme theory, the model defined by
the Lagrangian (\ref{Skymod}) reduces to the
Skyrme-Faddeev theory in the limit of constant
scalar field and $\xi=0$. So that, our model admits knot solutions
and a singular Wu-Yang monopole solution given by the hedgehog
configuration $\hat n=\dfrac{\vec r}{r}$.

Using stereographic projection it is convenient to introduce a complex
function $u(x)$ which parameterizes the two dimensional sphere $S^2\simeq CP^1$.
As usually, we assume that three-dimensional space $R^3$ is compactified
to a three-sphere $S^3$ by imposing appropriate boundary conditions at space infinity.
With this one has the following expression for the field
$\hat n$ in terms of the complex function $u(x)$
\bea
\hat n &=& \dfrac{1}{1+u \uast}
 \left (\ba{c}
  u+\uast\\
  -i (u-\uast)\\
 u \uast-1\\
            \ea
           \right ). \label{nstereo}
\eea
The magnetic field $H_{\mu\nu}$ can be written
explicitly in terms of $u$ as well
\bea
H_{\mu\nu}&=& \dfrac{-2 i}{(1+|u|^2)^2}
 (\der_\mu u \, \der_\nu \uast-\der_\nu u \,\der_\mu \uast).
 \eea
Since the magnetic field $H_{\mu\nu}$ represents
a closed differential two-form, one can find a
corresponding dual Abelian magnetic potential $\tC_\mu$
\bea
H_{\mu\nu}&=&\der_\mu \tC_\nu-\der_\nu \tC_\mu.
\eea
The Hopf and monopole charges for a given mapping $u(x)$
can be calculated as follows $(i,j=1,2,3)$
\bea
Q_H&=& \dfrac{1}{32 \pi^2}\int d^3x \epsilon^{ijk} \tC_i H_{jk}, \nn \\
g_m&=&\int_{S^2} H_{ij} d \sigma^{ij}.
\eea

Let us remind the main result obtained in study of
Aratyn-Ferreira-Zimerman integrable model \cite{AFZ}.
The Lagrangian of AFZ model is given by
\bea
{\cal L}_{AFZ}=-\eta_0 (H_{\mu\nu}^2)^{\frac{3}{4}}.
\eea
The model possesses a family of exact knot solutions
with a Hopf charge $Q_H=mn$ expressed in terms
of two winding numbers $(m,\,n)$.
For our further purpose
it is suitable to rewrite the analytic solutions
of AFZ model in spherical coordinates.
In the case of equalled winding numbers, $m=n$, the solution reads
 \bea
u&=&\dfrac{e^{i m \phi}}{2ar \sin \theta}(2ar \cos \theta-i(a^2-r^2)) \cdot \nn \\
&&\Bigg ( \dfrac{2ar \cos \theta-i(a^2-r^2)}{\sqrt {(a^2+r^2)^2-4 a^2 r^2 \sin^2 \theta}}\Bigg )^{m-1},
 \label{umm}
\eea
where $a$ is a free parameter characterizing the size of the knot.
In the case of $m\neq n$ an exact knot solution takes the form
\bea
u&=&e^{i n \phi}\Big (\dfrac{2ar \cos \theta-i(a^2-r^2)}{d}\Big )^m \cdot \nn \\
&&\Bigg (-\dfrac{a^2+r^2-\sqrt{\frac{n^2 d}{m^2}+4a^2r^2\sin^2\theta}}{a^2+r^2-
    \sqrt {d+\frac{4m^2a^2r^2 \sin^2\theta}{n^2}}} \Bigg)^{\frac{1}{2}}, \nn \\
d&\equiv&a^4+r^4+2 a^2 r^2 \cos(2 \theta). \label{umn}
\eea

Let us consider first a special case of a reduced Lagrangian
obtained from (\ref{Skymod}) by setting a condition $\beta=0$.
In this case one can find exact knot solutions with
Hopf charges determined by equalled winding numbers, $Q_H=m^2$.
By explicit calculating we have checked
for Hopf charges $Q_H$ corresponding to
winding numbers $m=n=1,2,...,6$  that field
configurations (\ref{umm}) and
the scalar field given by
\bea
\phi=\dfrac{c_0}{\sqrt {a^2+r^2}}, \label{solfi}
\eea
provide exact solutions to all equations of motion
under the following constraint
\bea
&&\mu^2- \dfrac{8 m^2(\nu+4 \xi)}{3a^2}=0. \label{constr1}
\eea
The integration constant $c_0$ in (\ref{solfi})
is fixed by boundary condition for $\phi$
at the origin. In analogy with the Skyrme theory
we will impose the following boundary
conditions for the scalar field
\bea
\phi(r=0)&=&1,~~~~~~\phi(r=\infty)= 0.
\eea
This implies $c_0=a$.
The constraint (\ref{constr1}) allows to determine the knot size
parameter
\bea
a_m^2=\dfrac{8 m^2(\nu+4 \xi)}{3\mu^2}.
\eea
A corresponding energy density of the solution is
\bea
{\cal E}_m=\dfrac{8 m^2(\nu+4 \xi) (r^2+3 a_m^2)}{3 (a_m^2+r^2)^3}.
\eea
Integrating the energy density over space
results in the following expression for
the total energy of the soliton
\bea
E_m=4\pi^2 \mu \, Q_H^{\frac{1}{2}} \sqrt {\dfrac{3}{8}(\nu+4\xi)} . \label{entot}
\eea
Notice, the energy of the knot is proportional to $Q_H^{\frac{1}{2}}$,
this is contrary to the usually expected dependence of
the low energy bound on the Hopf charge, $E_{bound}\simeq Q_H^{3/4}$,
found in a non-linear sigma model \cite{vakul}.
It would be interesting to find exact solutions for the Hopf
charge given by non-equalled winding numbers, $m \neq n$,
to verify whether the energy of knot solitons in that case
will have a total energy expressed by the same equation (\ref{entot}).

Let us consider a general case when the parameter $\beta$ does not vanish.
By direct calculating all equations of motion one can check that
an exact analytic solution with a unit Hopf charge, $Q_H=1$,
is given by the same expressions
(\ref{solfi},\, \ref{umm}) for $m=n=1$ and when the following
condition is fulfilled
\bea
\dfrac{3}{8} \mu^2+\beta-\dfrac{\nu+4 \xi}{a^2}=0.
\eea
If we keep the model parameters $\mu, \beta,\nu, \xi$
arbitrary the condition will fix the value of the knot size
\bea
a_1^2=\dfrac{\nu+4 \xi}{\dfrac{3}{8} \mu^2+\beta}.
\eea
It is easy to write down explicit expressions for the
components of the magnetic field
\bea
H_{r\theta}&=&-\dfrac{32 r^2 \sin \theta}{(a^2+r^2)^3}, \nn \\
H_{\theta \phi}&=& -\dfrac{8r^2\sin(2 \theta)}{(a^2+r^2)^2}, \nn \\
H_{r\phi}&=&-\dfrac{8r(a^2-r^2)(1-\cos(2\theta)}{(a^2+r^2)^3}.
\eea
Vector field lines of the magnetic field projected onto
the planes $XOZ$ and $XOY$ are depicted in
Figs. 1,2 (in cartesian coordinates $x,y,z$).
One can verify that magnetic fluxes through the circle
of radius "a" in the plane $z=0$ and through the half plane
$y>0$ are quantized with a factor $4\pi$
which is twice larger than a corresponding factor
of magnetic flux quantization in the Maxwell theory
\bea
&& \Phi_m=\int H_{r\phi} dr d \phi = 4 \pi m, \nn \\
&& \Phi_n=\int H_{r\theta} dr d \theta = 4 \pi n.
\eea
So, the Hopf charge $Q_H=mn$ provides
the linking number for the magnetic field which
has a helical toroidal structure.

One can calculate the energy density for the knot
with a unit Hopf charge
\bea
{\cal E}_1=\dfrac{a_1^2(8(\beta a_1^2+4 \xi+\nu)+\mu^2 r^2)}{(a_1^2+r^2)^3}.
\eea
This implies the total energy
\bea
E_1=4 \pi^2 \sqrt{(\beta+\frac{3}{8}\mu^2)(\nu+4 \xi)}.
\eea

Surprisingly, as it was observed first in \cite{nicole},
the knot configuration has a very nice feature.
Even though the magnetic field possesses a toroidal topology,
the energy density is completely spherically symmetric
(see Fig. 2a).

\begin{figure}[htp]
\centering
\subfigure[~]{\includegraphics[width=40mm,height=40mm]{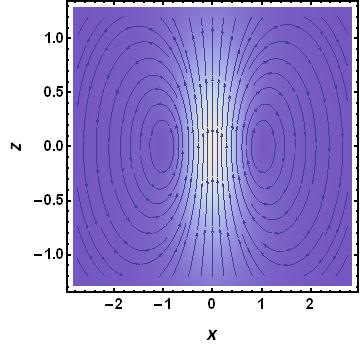}}
\hfill
\subfigure[~]{\includegraphics[width=40mm,height=40mm]{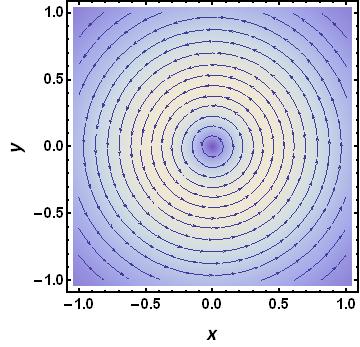}}\\
\caption[Hmnplots]{Vector lines of the magnetic field projected onto
the planes $y=0$, (a), and $z=0$, (b).}\label{fig:Hmn}
\end{figure}

For higher Hopf charges, $Q_H > 1$,
numeric analysis implies
that solution for the scalar field  $\phi$
is not spherically-symmetric.
To find approximate knot solutions with higher Hopf numbers
we apply minimization procedure for the energy functional.
A key point is how to find
a proper ansatz for the function $u(x)$
which could provide a minimum energy for
a chosen Hopf charge.
We apply first the ansatz for $u(x)$
given by AFZ configurations (\ref{umn}).
To find a dominant contribution
to the energy we introduce the following ansatz for
the scalar field $\phi$
\bea
&& \phi=\dfrac{c_1}{\sqrt{c_2^2+r^2}}+\dfrac{c_3}{\sqrt{c_4^2+r^2}} \sin \theta,
\eea
where $c_1,c_2,c_3,c_4$ are variational number parameters.
We set the values of the model parameters in the Lagrangian
(\ref{Skymod}) to
$(\xi=0, \beta=0.33,\mu=1, \nu=1.5)$.
Without loss of generality we put $\xi=0$ since the presence of the
last term in (\ref{Skymod}) does not change a qualitative picture
of the solution structure.
Minimization of the energy with the ansatz based on AFZ solutions
leads to the following estimates for the total energy and
knot size parameter $a$ given in Table I. One can see that
dependence of the knot energy on Hopf charge is in a good agreement
with the low energy bound found in \cite{vakul}
\bea
&& E_{bound} = E_1 Q_H^{3/4}. \label{enbound}
\eea
A significant discrepancy appears for knots with Hopf charge
defined by equalled winding numbers, $Q_H=m^2$.
For instance, for Hopf charge $Q_H=4,~m=n=2$
the energy minimization procedure
with the ansatz (\ref{umm},\ref{solfi})
produces energy value $E_4 (m=n=2)=105.7$
which is less than a respective low energy bound.
This is caused by the essential feature of the AFZ
configurations, namely, by the
spherical symmetry of the
energy densities for $m=n=1,2$ (see Figs. 2a, 2d).

Let us now consider the ansatz with rational functions
$u(z_1,z_2)$ which was successfully applied in the
Skyrme-Faddeev model \cite{battye2013}.
The ansatz starts with the definition of
mapping $R^3 \rightarrow SU(2)\simeq S^3$
\bea
z_1&=&\dfrac{x_1+i x_2}{r} \sin f(r), \nn \\
z_2&=&\cos f(r)+i \dfrac{x_3}{r}\sin f(r),
\eea
where $f(r)$ is a trial function,
 and the complex coordinates $z_1,z_2$
describe the sphere $S^3$ imbedded into the
complex plane $C^2$
\bea
|z_1|^2+|z_2|^2=1.
\eea
A class of toroidal knot configurations of type
${\cal A}_{m,n}$ with a Hopf charge $Q_H=mn$
 is defined by the rational function \cite{battye2013}
\bea
u=\dfrac{z_2^m}{z_1^n}.
\eea
We choose a following profile function $f(r)$ in the interval $(r_0=0.01, r_f=100)$
satisfying the boundary conditions $f(r_0)=\pi$ and $f(r_f)=0$
\bea
f(r)&=&\dfrac{\pi (r_f-r)(1+b_0 (r-r_0)}{r_f-r_0}e^{-b_1^2 (r-r_0)},
\eea
where $b_0,b_1$ are variational parameters.
The profile function for the scalar field is
defined in the form
\bea
g(r,\theta)&=&\Big (\dfrac{c_0}{\sqrt{a_0^2+r^2}}+\dfrac{c_1 r }{c_2^2+r^2}\sin \theta \Big)
 \dfrac{r_f-r}{r_f+r},
\eea
where $c_0,a_0,c_1,c_2$ are variational parameters,
and we keep only a leading term with angle dependence.
The boundary conditions for $g(r,t)$ in the interval $(r_0,r_f)$
are fixed by the requirement of finiteness of the total energy
\bea
g(0,\theta)&=& const,~~~~~
g(\infty, \theta)\simeq \Big (\dfrac{1}{r}\Big )^{\alpha \geq 1}.
\eea

We minimize the energy functional and
find the profile functions $f(r), g(r,\theta)$.
Respective energy density plots are depicted
in Fig. 3. The energy values of knots are presented in Table 1,
the last column $E_{rat}$.

\begin{table}
\begin{tabular}{| l | c | c | c | c | c |}
   \hline
  ~ (m,n) & $Q_H$ & $E_{bound}$ & $E_{AFZ}$ & $a_{AFZ}~$ & $E_{rat}~$ \\ \hline
  ~ (1,1) & 1 & 40.6 & 40.6  &~ 1.46~ &~ 41.6  ~\\ \hline
  ~ (1,2) & 2 & 68.3 & 67.28 & ~2.0  ~&~ 68.3  ~\\ \hline
  ~ (2,1) & 2 & 68.3 & 67.26 &~ 2.03~ &~ 75.0 ~ \\ \hline
  ~ (1,3) & 3 & 92.5 & 96.9  & ~1.53~ &~ 107.8~ \\ \hline
  ~ (3,1) & 3 & 92.5 & 98.8  & ~0.51~ & ~116 ~ \\ \hline
  ~ (1,4) & 4 & 115  & 129   &~ 1.49 ~& ~149 ~ \\ \hline
  ~ (4,1) & 4 & 115  & 133   & ~0.50~ &~ 164 ~ \\ \hline
  ~ (2,2) & 4 & 115  & 105.7 & ~0.57~ &~ 123 ~ \\
   \hline
 \end{tabular}
\caption{Energy and knot size values for Hopf charge $Q_H=mn$.
Values for the model parameters are $\beta=0.33,~\mu=1,~\nu=1.5,~\xi=0$. }
\end{table}

Surprisingly, comparison of the results obtained by using
energy minimization with the rational profile functions with the
results obtained by using AFZ configurations implies that
exact knot solutions of AFZ integrable model
provide good approximate analytic solutions for our model.
Notice, numeric analysis demonstrates that one has an
essential qualitative difference between
approximate variational solutions obtained with AFZ field configurations
and with rational functions, (see Figs. 2,3).
In the case of use of the rational function
ansatz the energy densities for any Hopf charge
are not spherically symmetric.
\begin{figure}[htp]
\centering
\subfigure[~$m=n=1$ ]{\includegraphics[width=38mm,height=36mm]{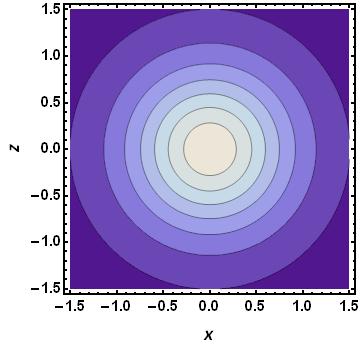}}
\hfill
\subfigure[~$m=1,n=2$ ]{\includegraphics[width=38mm,height=36mm]{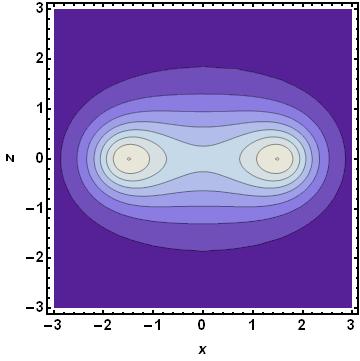}}\\
\subfigure[~$m=2,n=1$ ]{\includegraphics[width=38mm,height=36mm]{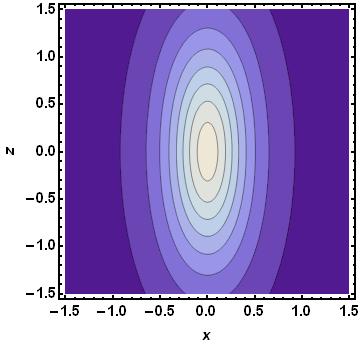}}
\hfill
\subfigure[~$m=n=2$ ]{\includegraphics[width=38mm,height=36mm]{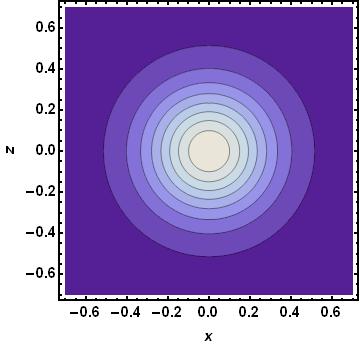}}\\
\caption[AFZplots]{Energy density plots
with AFZ ansatz for Hopf charge $Q_H=mn$}\label{fig:AFZ}
\end{figure}
\begin{figure}[htp]
\centering
\subfigure[~$m=n=1$ ]{\includegraphics[width=38mm,height=36mm]{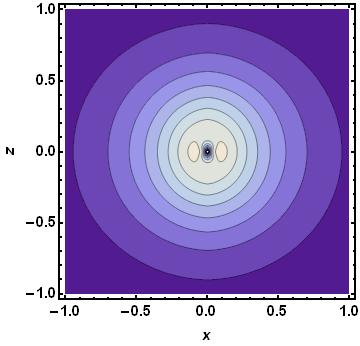}}
\hfill
\subfigure[~$m=1,n=2$ ]{\includegraphics[width=38mm,height=36mm]{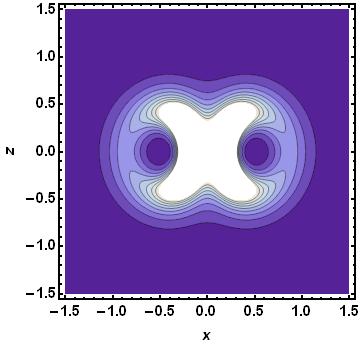}}\\
\subfigure[~$m=2,n=1$ ]{\includegraphics[width=38mm,height=36mm]{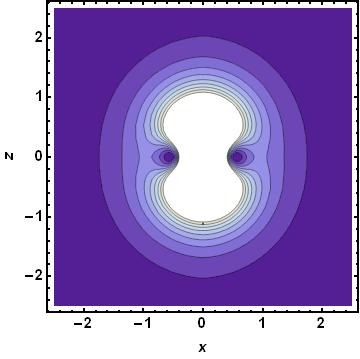}}
\hfill
\subfigure[~$m=n=2$ ]{\includegraphics[width=38mm,height=36mm]{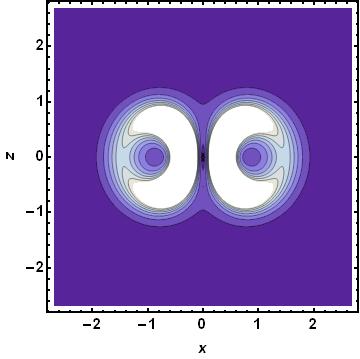}}\\
\caption[RATplots]{Energy density plots
with rational functions
ansatz for Hopf charge $Q_H=mn$}\label{fig:RAT}
\end{figure}
Notice, the Lagrangian (\ref{Skymod}) has another interesting
limiting case with a constrained set of the model parameters.
When the parameters $\nu=\xi=0$ one has a special model without
any dimensional parameters while possessing
a static solution given by (\ref{umm}, \ref{solfi}), $Q_H=1$,
with an additional constraint
\bea
\dfrac{3}{8} \mu^2+\beta=0.
\eea
The parameter $a$ remains a free parameter.
The solution has a spherically symmetric
energy density
\bea
{\cal E}_0= \dfrac{a^2 \mu^2 (r^2-3 a^2)}{(a^2+r^2)^3}.
\eea
One can easily verify that a total energy vanishes identically,
so the solution represents a non-trivial zero mode.
A similar solution exists in a special critical case
in an extended Skyrme-Faddeev model
considered in \cite{ferr2004}. It would be interesting
to study possible physical implications of such models
especially in description of monopoles.

In conclusion, we have proposed a generalized Skyrme-Faddeev
model with an additional scalar field. The model admits an
exact knot soliton with a unit Hopf charge and inherits main
features of the Aratyn-Ferreira-Zimerman  integrable $CP^1$
model. Namely, our numeric analysis shows that AFZ knot configurations
provide approximate analytic knot solutions with high
symmetric properties.
Notice, that the $CP^1$ field $\hat n$ realizes another
non-trivial topological mappings classified by
the second homotopy group $\pi_2(CP^1)=Z$, i.e.,
by the monopole charge. Since the model admits the
Wu-Yang monopole as a singular solution, and there is an
additional scalar field $\phi$ which can regularize
the solution structure, it would be
interesting to study possible finite energy monopole solutions.
As it was suggested before, the Skyrme-Faddeev theory and
its various extensions can represent low energy
effective theories of QCD. Construction of a consistent
extended Skyrme-Faddeev model for description of mesons and
glueballs is an important issue in modern hadron spectroscopy.
\acknowledgments
The work is supported by
NSFC (Grants 11035006 and 11175215), CAS (Contract No. 2011T1J31),
and by UzFFR (Grant F2-FA-F116).

\end{document}